%% file: daddio.tex
\documentclass[graybox]{svmult}

\usepackage{mathptmx}       
\usepackage{helvet}         
\usepackage{courier}        
\usepackage{type1cm}        
\usepackage{makeidx}         
\usepackage{graphicx}        
\usepackage{multicol}        
\usepackage[bottom]{footmisc}

\usepackage{amsmath}
\usepackage{latexsym}
\usepackage{amsfonts}
\usepackage{amssymb}
\usepackage[dvips]{epsfig}
\usepackage{psfrag}
\usepackage{geometry}
\usepackage{graphicx}
\usepackage{floatrow}
\usepackage{subfig}
\usepackage[english]{babel}
\usepackage{times}
\usepackage[utf8x]{inputenc}
\usepackage{pgf}
\usepackage{color}
\usepackage{multimedia}
\usepackage{xcolor}
\usepackage[thicklines]{cancel}
\usepackage{setspace}
\usepackage{mathrsfs}
\usepackage{txfonts}

\makeindex             

\begin{document}

\title*{Dynamics of the free jets from nozzles of complex geometries}
\author{Paolo D'Addio and Paolo Orlandi}
\institute{Paolo D'Addio \at Dipartimento di Ingegneria Meccanica e 
Aerospaziale, ”Sapienza” Università di Roma, Italy, 
\email{paolo.daddio@uniroma1.it}
\and Paolo Orlandi \at Dipartimento di Ingegneria Meccanica e Aerospaziale,
”Sapienza” Università di Roma, Italy}
%
%
\maketitle

\vspace{-1.8cm}
\section*{Abstract}
The dynamics of the coherent structures in jets generated by nozzles of
different shapes is analyzed through DNS at $Re_{D_e}=565$, by considering
circular, square, fractal and star-like nozzles. The jets generated from 
orifices with corners, undergone a rotation proportional to the corner 
angular width: $\theta_{rotation}=\theta_{corner}/2$. The velocity at which
this rotation occurs is also affected by the angle of the corners, being 
faster for fractal and star-like nozzles which have small 
$\theta_{corner}$. Therefore it has been found that the velocity of the 
rotation is associated with enhanced spreading and entraining 
characteristics. The jet evolution and its rotation are dictated by the 
vorticity field and, in particular, by the positive and negative $\omega_x$
layers generated at each corner. The comparison between the fractal and the
star-like jets at this $Re$, suggested that the effect of the smaller 
scales generated by the fractal nozzle does not play a role in the 
development of the jet, that evolves as the star-like one.

\vspace{-0.2cm}
\section{Introduction}
The starting jet formed by the sudden discharge of hot reaction products 
into a fuel-air mixture may have a large influence on the combustion 
process \cite{w1985}, with important implications in the transport, 
handling and storage of fuels, particularly hydrogen, and may also find 
application in engine ignition systems \cite{m1996}. The control of the jet
evolution is strongly dependent on understanding the dynamics of the 
vortical structures, because the spreading is affected by the formation, 
interaction, merging, and breakdown of these structures. 

The starting jet generated by a circular orifice shows a leading vortex 
ring followed by a slender jet stem, and the associated flow dynamics has 
been studied for constant density configurations by \cite{g1998}. They 
observed that, as the boundary layer separates at the orifice, the vortex 
sheet rolls up to form a toroidal vortex that travels downstream, 
entraining the outer fluid. As a result of the roll-up, a mixed core of 
reactants and combustion products should form at the jet head, providing a 
precursor ignition kernel where chemical reactions are enabled by the high 
temperature. Successively, \cite{hh1983} and \cite{hg1987} found that the 
entrainment of jets can be enhanced by using non-circular nozzles. The 
reason for the enhanced mixing and entrainment properties has been 
attributed to the characteristic rotation of the jet cross section, that 
occurs during its development in the stream-wise direction. This axis 
rotation results from self-induced deformation of vortex rings with 
non-uniform azimuthal curvature, and it has been observed using elliptic 
nozzles and nozzles with corners (\cite{gsphw1989}, \cite{th1989}). In 
particular, laboratory experiments by \cite{gsphw1989} have shown that such
rotation in square jets is responsible for enhancing the mixing of fluid in
the neighborhood of the corner regions and near the jet outlet.

In the present work, DNS simulations at low Reynolds number have been 
performed to study the near-field evolution of the vortical structures
generated by nozzles of complex shapes. 

\vspace{-0.2cm}
\section{Numerical experiments}
\begin{figure}[h!]
\begin{center}
\sidesubfloat[]
{\includegraphics[width=0.2\textwidth]{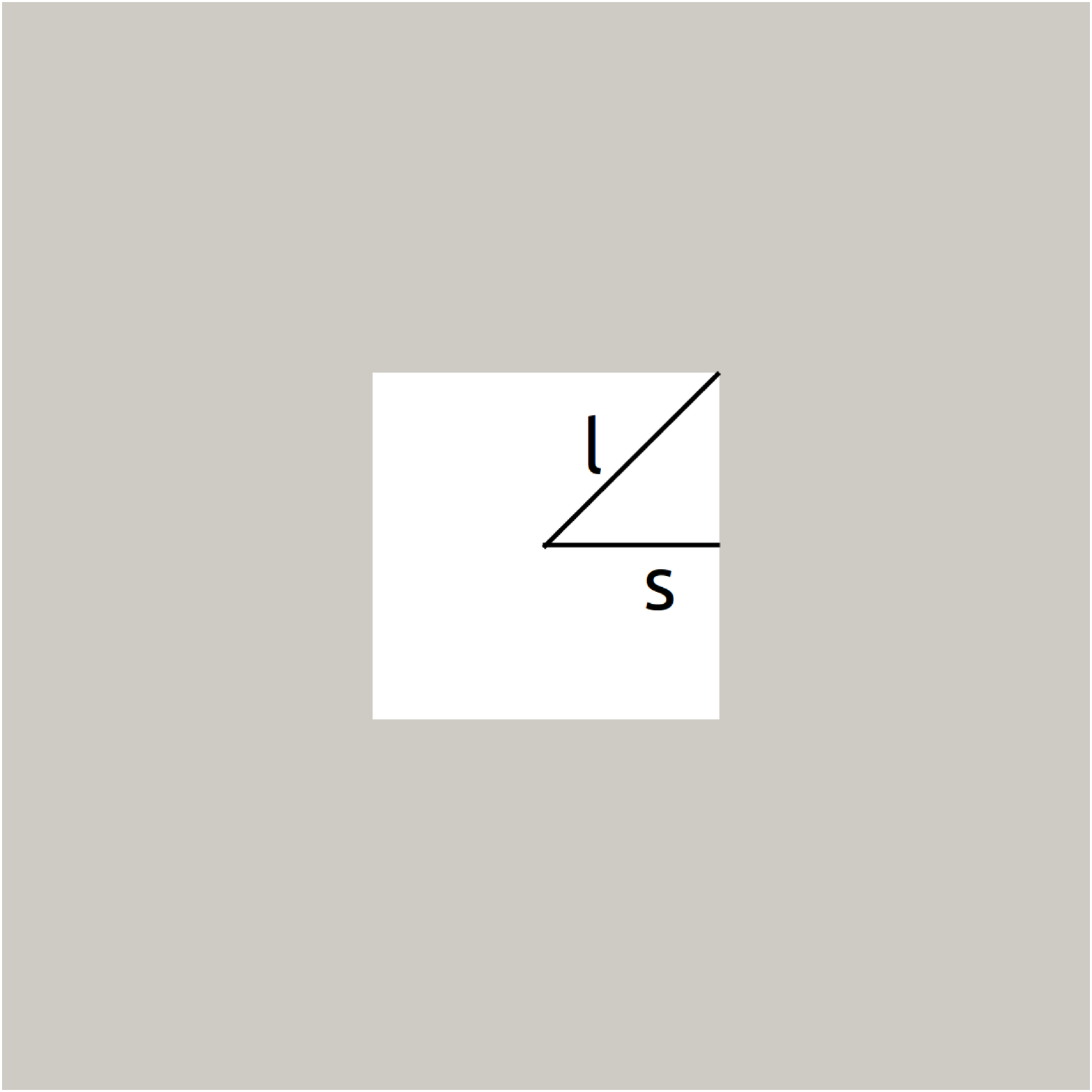}}
\sidesubfloat[]
{\includegraphics[width=0.2\textwidth]{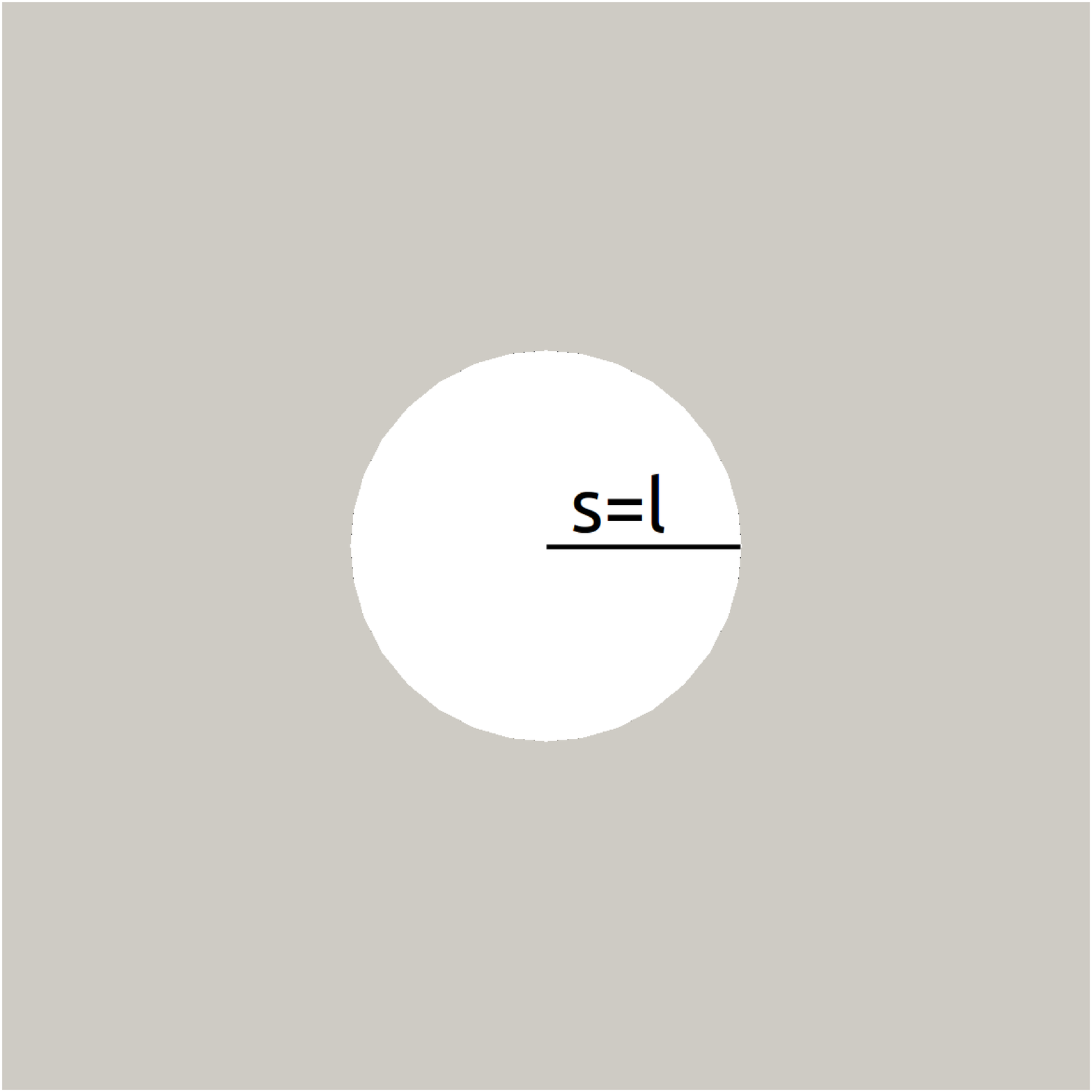}}
\sidesubfloat[]
{\includegraphics[width=0.2\textwidth]{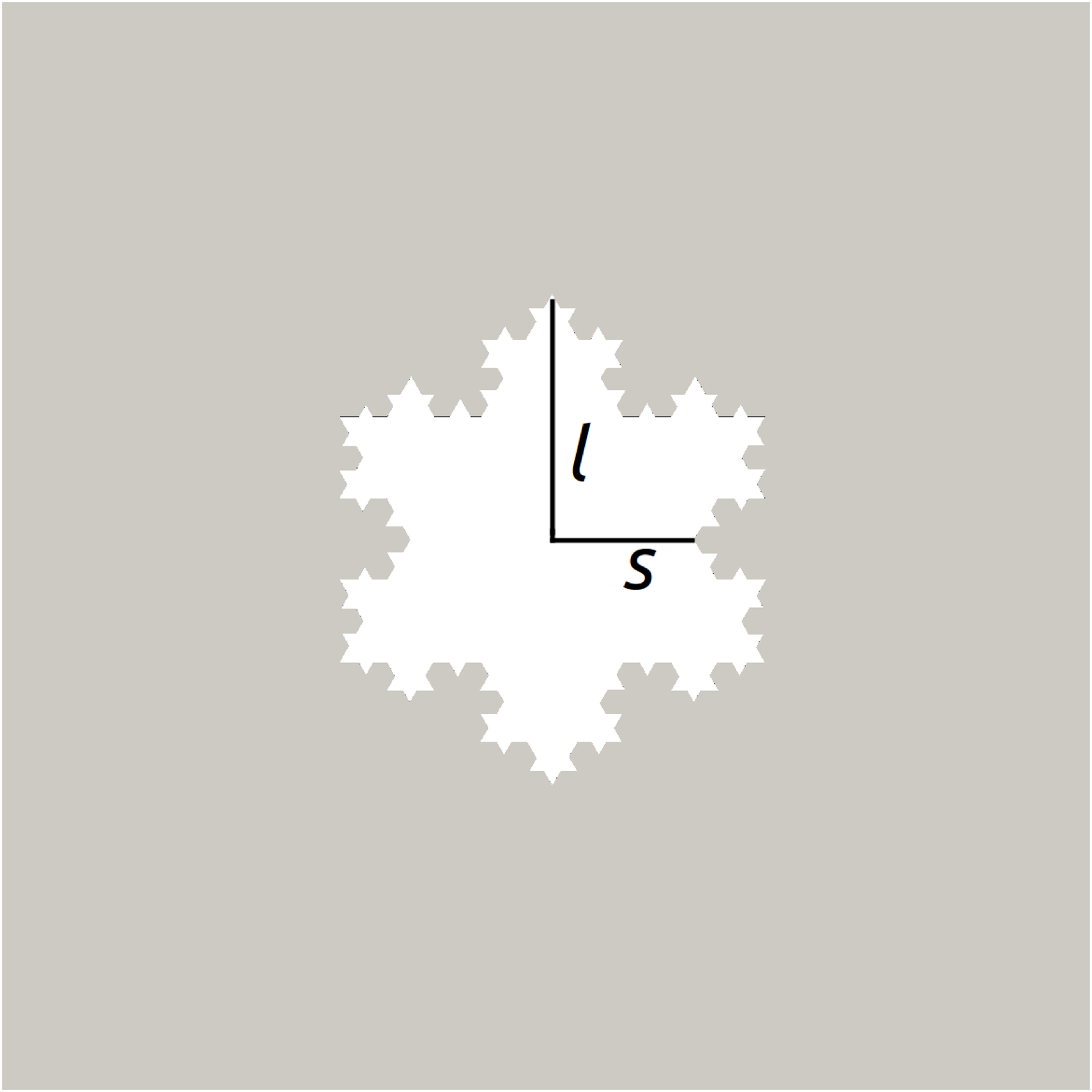}}
\sidesubfloat[]
{\includegraphics[width=0.2\textwidth]{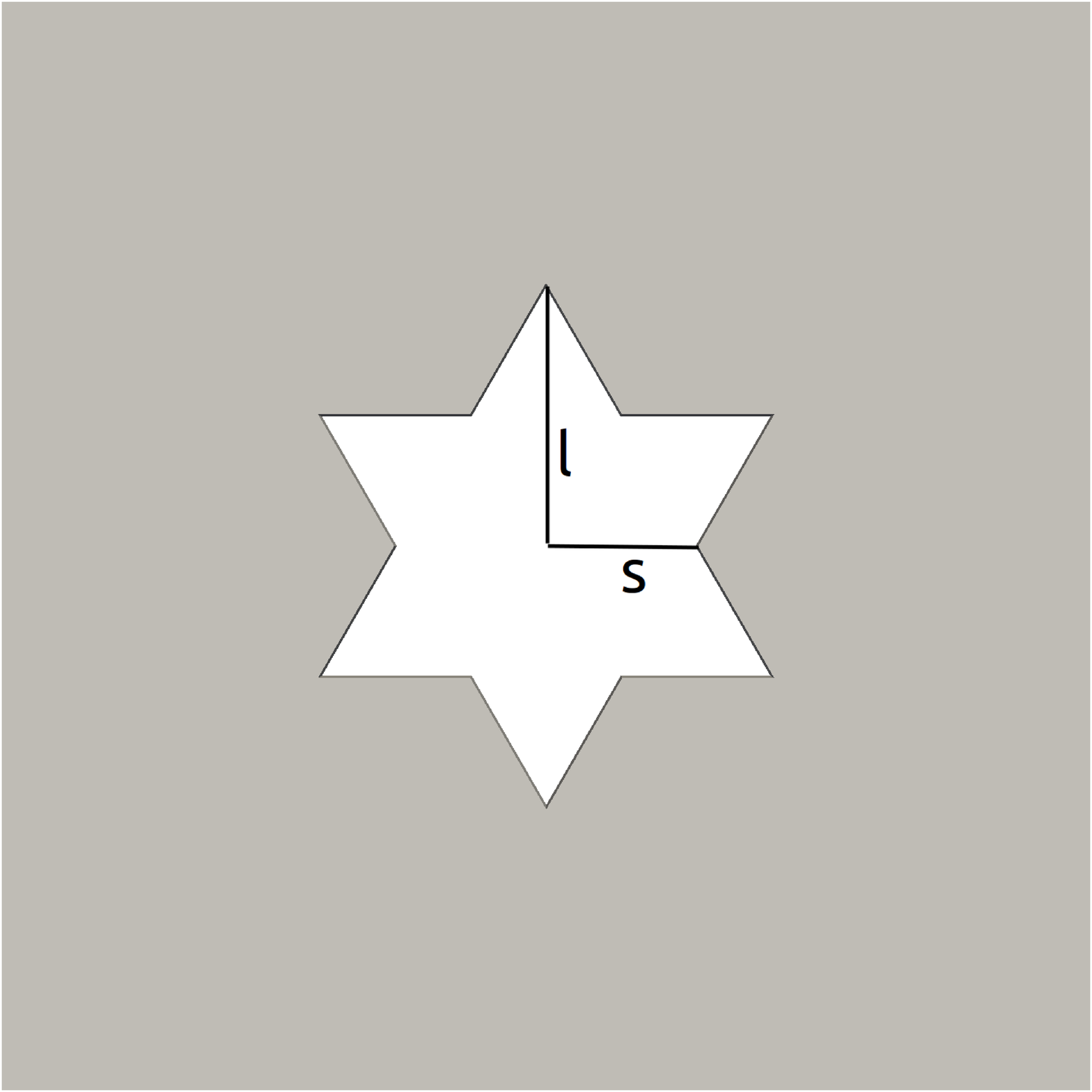}}\\
\caption{Geometry of the orifices $(a)$ $J_S$, $(b)$ $J_C$, $(c)$ $J_F$,
$(d)$ $J_6$ and $s$- and $l$-directions.}
\label{geoJsl}
\end{center}
\end{figure}

The details of the numerical scheme to solve the Navier-Stokes equations, 
together with the immersed boundary method used to reproduce the 
interaction between the flow and the solid, are described in \cite{ol2006}.
The shape of the orifices considered are given in figure \ref{geoJsl}. The 
fractal orifice has been obtained with three iterations of the basic 
triangular geometry and all the orifices have been designed to have the 
same area, thus the solidity is the same in all the cases. $D_e$ is the 
equivalent diameter of the jets and it is $D_e=2.257$ since the area is 
equal to $4$. At the inlet an uniform velocity profile is imposed and the 
orifice is located at $x_N=2.25$. The simulations were performed at 
$Re_{D_e}=U_{\infty}D_{e}/\nu=565$ ($U_{\infty}=1$) in a computational 
domain $L_1 \times L_2 \times L_3=3\pi\times 2\pi \times 2\pi$ discretized 
with $n_1 \times n_2 \times n_3=385\times 257 \times 257$ points. The 
orifice is located in $y-z$ and $x$ is the downstream direction. 
Periodicity is assumed in $y$ and $z$. In the plots the $x$-coordinate is 
normalized with $D_e$ starting from $x_N$: $x^*=(x-x_N)/D_e$.
\floatsetup[figure]{style=plain,subcapbesideposition=top}
\vspace{-0.2cm}
\section{Results}
\begin{figure}[h!]
\begin{center}
\sidesubfloat[]
{\includegraphics[width=0.35\textwidth,angle=0]{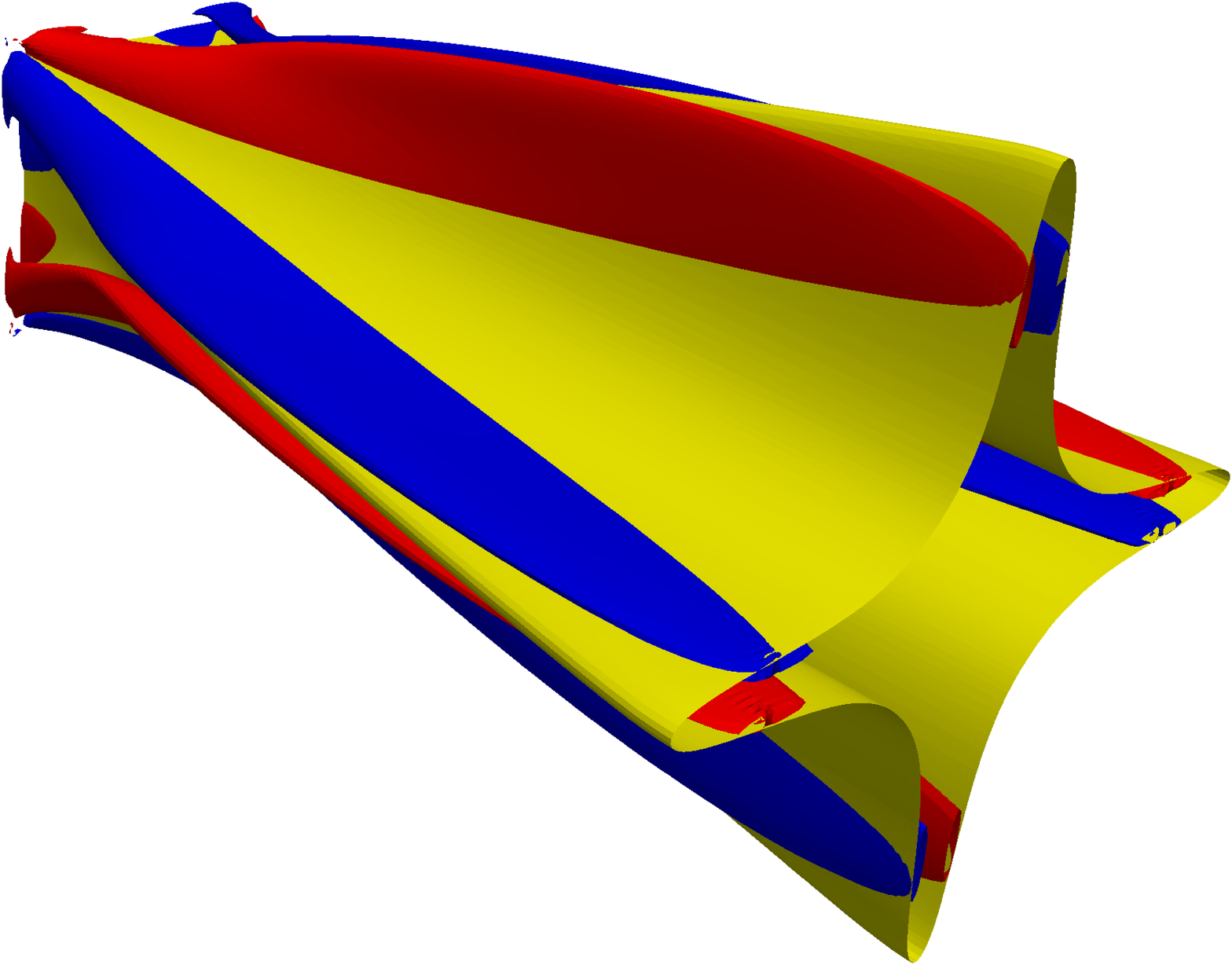}}
\sidesubfloat[]
{\includegraphics[width=0.35\textwidth,angle=0]{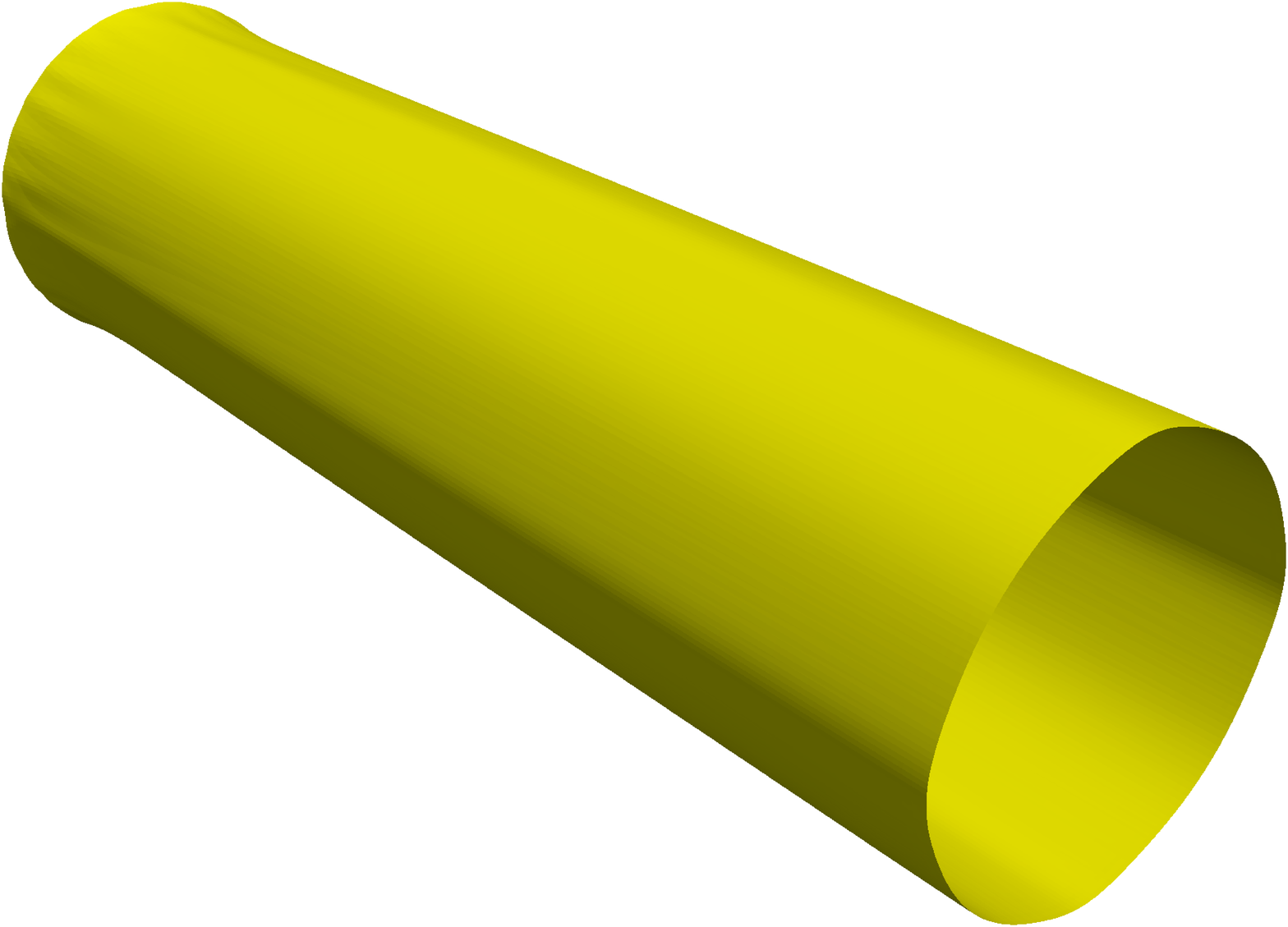}}\\
\sidesubfloat[]
{\includegraphics[width=0.35\textwidth,angle=0]{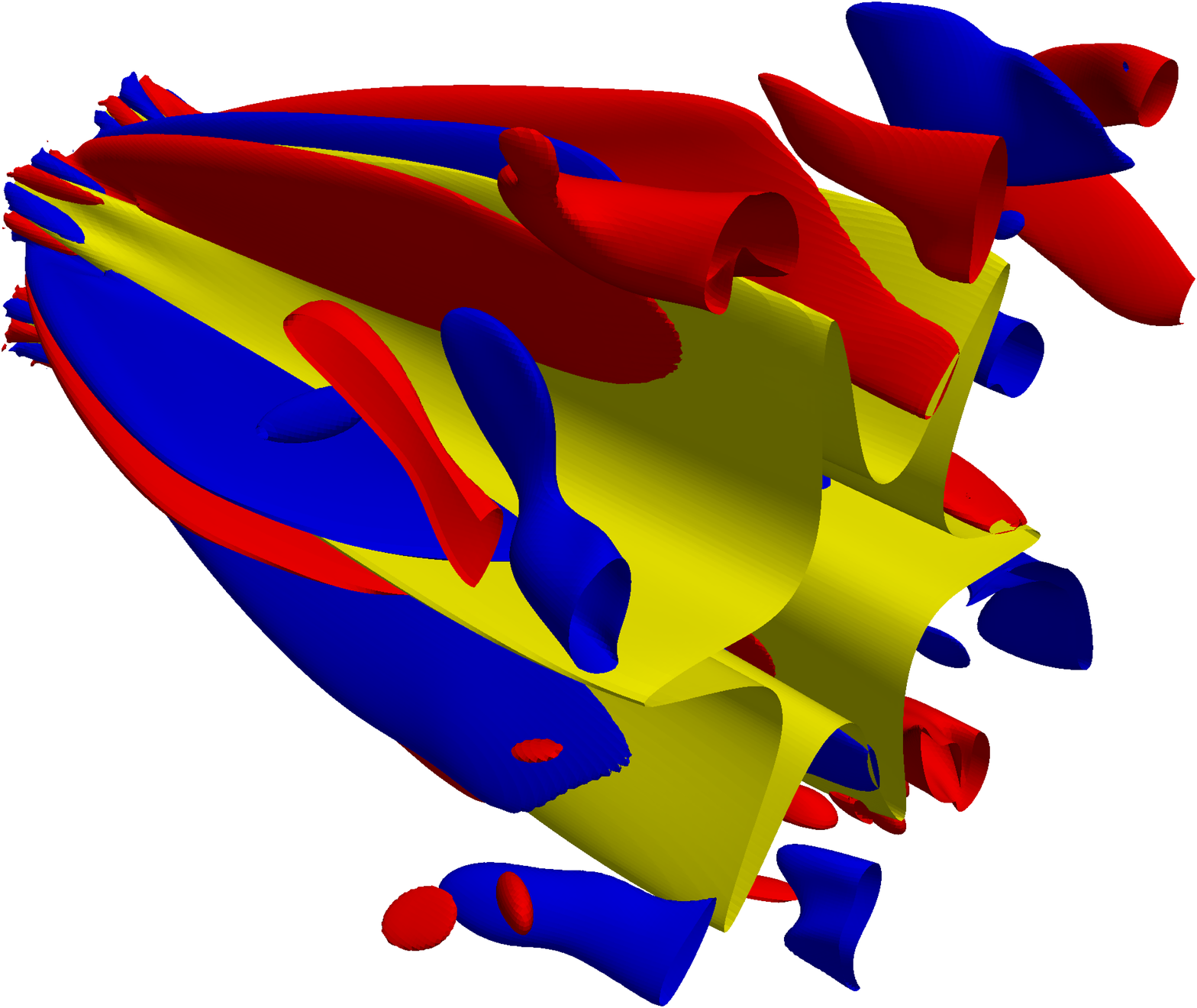}}
\sidesubfloat[]
{\includegraphics[width=0.35\textwidth,angle=0]{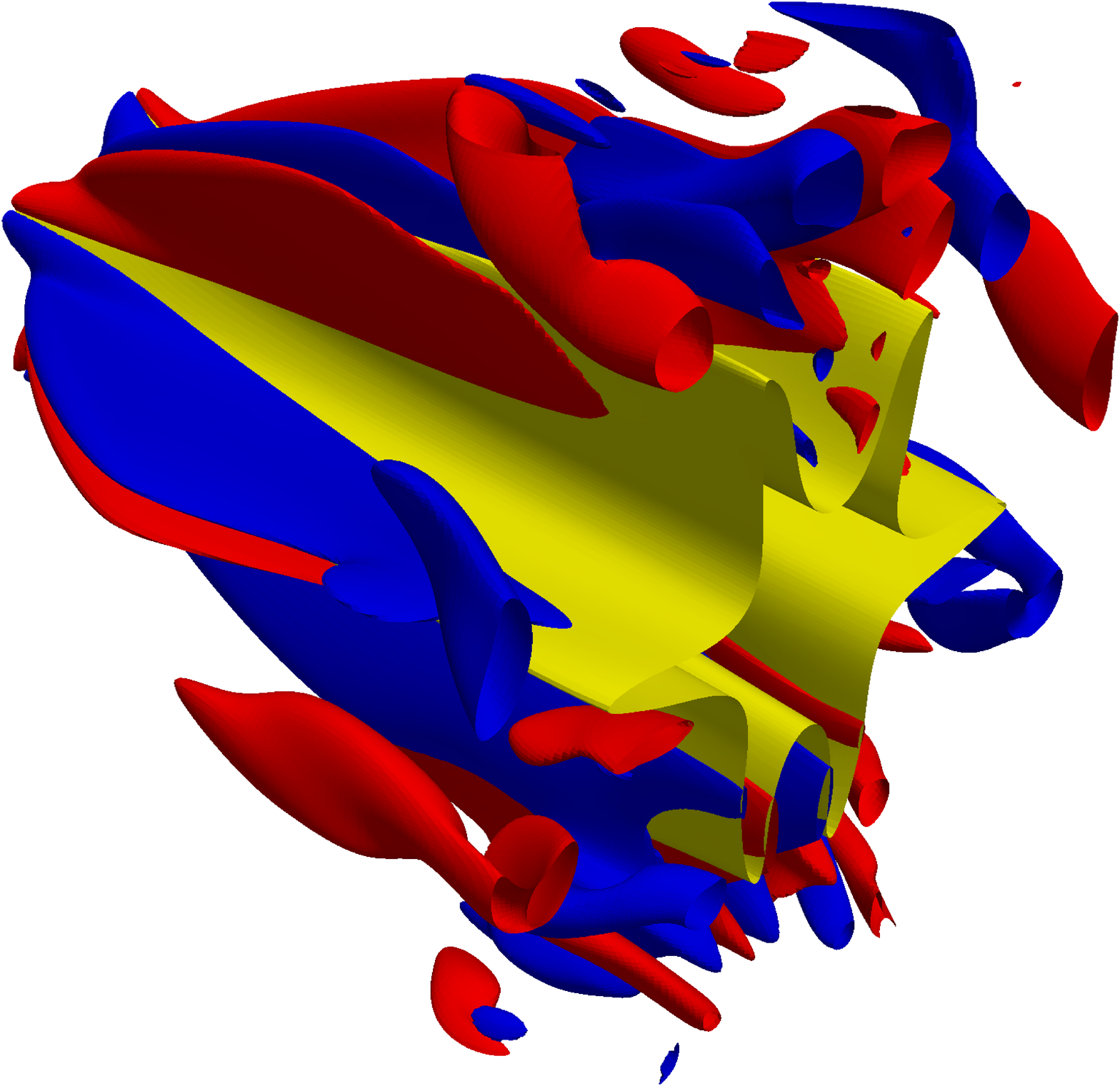}}\\
\caption{3D Contour plot of $U=U_{\infty}/2$ for $(a)$ $J_{S}$, $(b)$ 
$J_{C}$, $(c)$ $J_{F}$ and $(d)$ $J_{6}$ in yellow, overlapped with 
$\omega_x/\omega^{max}= \pm 0.5$ in red and blue respectively. 
$0<x^*<3.18$.}
\label{OXV_J}
\end{center}
\end{figure}

These DNS allow to study the vortex dynamics governing the evolution of the
jets. At $Re=565$, the flow is laminar at the exit from the nozzle, and 
therefore it is possible to understand the causes leading to the jet 
deformation. In figure \ref{OXV_J} the surface contour of $U=U_{\infty}/2$ 
are given for the square, circular, fractal and star-like jets in yellow. 
The flow generated by the circular orifice ($J_C$) is spread by viscous 
diffusion and maintains its circular shape downstream; in the other cases 
the shape is dictated by the hole only in the proximity of the orifice, and
is lost is a short distance. At the end of the domain the flow is 
characterized by a number of corrugations that are rotated with respect to 
their initial position. For $J_S$ these are rotated of $45^{\circ}$; 
for $J_F$ only the six corrugations, related to the largest scales, are 
recognizable from the contour of $U$, and these are rotated about 
$30^{\circ}$ with respect to the initial configuration. The same rotation 
occurs for $J_6$. Therefore in the cases with corners, the rotation of the 
jet is proportional to the number of spikes, in particular: 
$\theta_{rotation}=\theta_{corner}/2$. This rotation was studied 
experimentally and numerically by \cite{ggp1995} and \cite{mmg1995}; in 
particular \cite{mmg1995} considered square, rectangular and equilateral 
triangular orifices, observing a $45^{\circ}$ rotation for the first, an 
axis-switching ($90^{\circ}$ rotation) for the second, and an overturning 
of $180^{\circ}$ for the latest. Considering only the initial and final 
position, the overturning of the equilateral triangle is indeed equivalent 
to a $30^{\circ}$ rotation, with $\theta_{rotation}=\theta_{corner}/2$.
Therefore also the work of \cite{mmg1995} corroborates the present results 
of $\theta_{rotation}=\theta_{corner}/2$, which is valid only for the 
shapes with aspect ratio equal to $1$. In fact the rectangular and the 
elliptic orifice, undergo a different $\theta_{rotation}$.
\floatsetup[figure]{style=plain,subcapbesideposition=top}

\begin{figure}[h!]
\begin{center}
\sidesubfloat[]
{\psfrag{X}[][][1]{$x^*$}
\psfrag{Y}[][][1]{$r_{1/2}/D_e$}
\psfrag{T}[][][1]{}
\includegraphics[width=0.3\textwidth,angle=270]{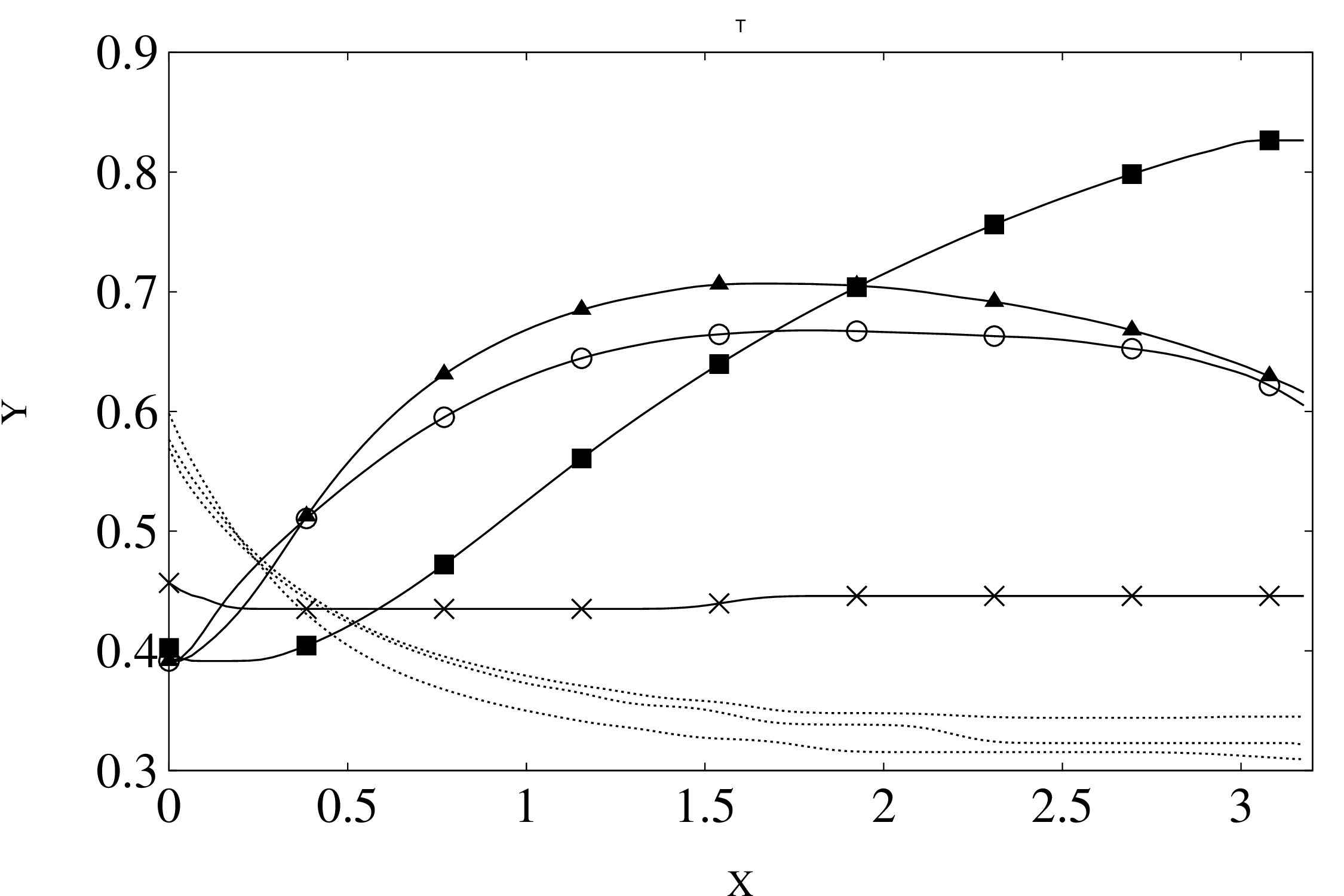}}
\sidesubfloat[]
{\psfrag{X}[][][1]{$x^*$}
\psfrag{Y}[][][1]{$\left\langle \Omega_{\theta} \right\rangle$, 
$\left\langle \Omega_x \right\rangle$}
\psfrag{T}[][][1]{}
\includegraphics[width=0.3\textwidth,angle=270]{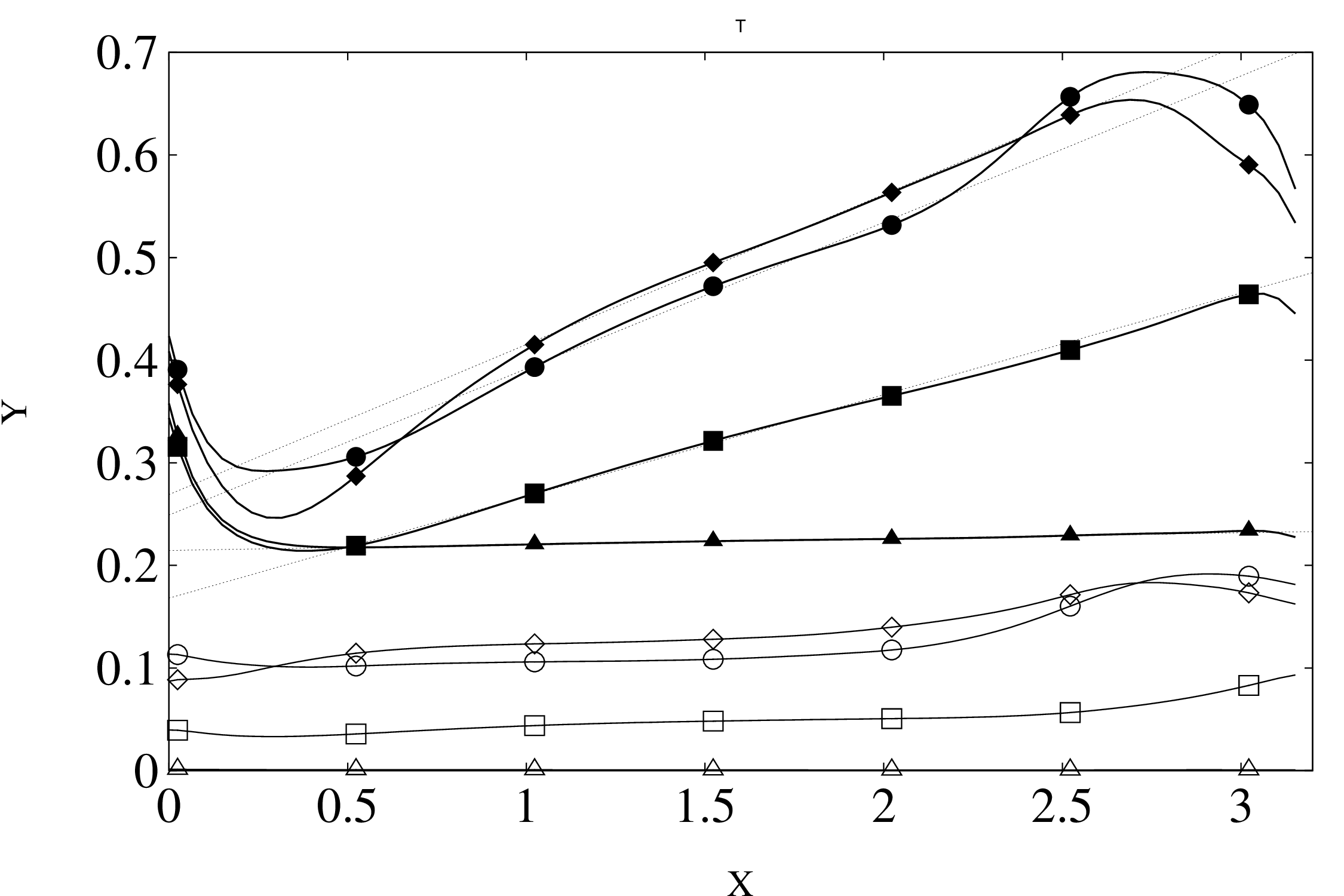}}
\caption{$(a)$ Evolution of the jet half-width normalized with $D_e$ vs 
stream-wise direction $x^*$. Symbols: ($\blacksquare$): $r_{1/2}^s$ for 
$J_S$; ({\color{black} $\circ$}): $r_{1/2}^s$ for $J_F$;
({\color{black} $\blacktriangle$}): $r_{1/2}^s$ for $J_6$;
({\color{black} $\times$}): $r_{1/2}$ for $J_C$. Dashed lines for the
$r_{1/2}^l$. $(b)$ Evolution of 
$\left\langle \Omega_{\theta} \right\rangle$ 
and $\left\langle \Omega_x \right\rangle$ vs the stream-wise 
direction $x^*$. Symbols: ($\square$): 
$\left\langle \Omega_x \right\rangle$ for $J_S$; ($\blacksquare$):
$\left\langle \Omega_{\theta} \right\rangle$  for $J_S$; ({\color{black} 
$\circ$}): $\left\langle \Omega_x \right\rangle$ for $J_F$; 
({\color{black} $\bullet$}): $\left\langle \Omega_{\theta} \right\rangle$ 
for $J_F$; ({\color{black} $\Diamond$}): 
$\left\langle \Omega_x \right\rangle$ for $J_6$;({\color{black} 
$\Diamondblack$}): $\left\langle \Omega_{\theta} \right\rangle$ for $J_6$;
({\color{black} $\triangle$}): 
$\left\langle \Omega_x \right\rangle$ for $J_C$; 
({\color{black} $\blacktriangle$}): 
$\left\langle \Omega_{\theta} \right\rangle$ for $J_C$.}
\label{bhsf}
\end{center}
\end{figure}

The location where the rotation of the jet terminates is also affected by 
the shape, and this length can be determined by monitoring the stream-wise 
variation of the half velocity width $r_{1/2}$ measured on the smallest
($s$-direction) and largest ($l$-direction) radii defined in figure 
\ref{geoJsl}. In figure \ref{bhsf}a we reported the variation of the 
half-widths $r_{1/2}^s$ and $r_{1/2}^l$ non-dimensionalized by $D_e$ for 
$J_S$, $J_F$ and $J_6$ while for $J_C$ we reported only $r_{1/2}^s$ because
the jet remains circular. The half-width $r_{1/2}$ of the jet, at the given
stream-wise location, is the distance from the center-line at which the 
axial velocity drops to half of its center-line value $U_{cl}$. If 
$r_{1/2}^s$ becomes larger than $r_{1/2}^l$ means that the jet has inverted
its long and short diameters, and the rotation 
$\theta_{rotation}=\theta_{corner}/2$ is completed. $r_{1/2}^s$ becomes 
larger than $r_{1/2}^l$ at $x^*=0.513$ for $J_S$, $x^*=0.260$ for $J_F$ and
$x^*=0.275$ for $J_6$, thus $J_F$ and $J_6$ complete their rotation before 
$J_S$. $r_{1/2}^s-r_{1/2}^l$ is also a fundamental quantity because 
indicates the stretching of the corner portions of the jet and, being 
$r_{1/2}^l$ approximately equal for all the jets, where $r_{1/2}^s$ grows 
faster, the corners undergo a stronger stretching. After the initial region
where $J_F$ is spread faster than $J_S$, at $x^*\approx 1.5$, $r_{1/2}^s$ 
reaches its maximum, while it continues to grow for $J_S$. At the end of 
the domain $r_{1/2}^s$ of $J_S$ is larger than in $J_F$ because the former 
maintains its coherent shape longer, and the spreading is slower; on the
other hand $J_F$, as it will be shown later, produces an intricate flow 
pattern, where the vortex interactions make the jet to lose its coherent 
shape, and $r_{1/2}^s$ decreases.

To understand whether small scales corrugations play a role at small $Re$
the fractal jet has been compared with $J_6$. Figure \ref{bhsf}a does not 
show large differences, suggesting that the smallest scales generated by 
the fractal jet die and do not influence the evolution far from the nozzle.
The simulation $J_6$ allows to attribute the initial highest spreading of 
$J_F$, not to the fractal corrugations, but to the angular width of the 
largest corners. In figure \ref{bhsf}a it is reported also $r_{1/2}$ for 
$J_C$ to show that with this Reynolds number the radius of the jet remains 
almost unchanged along $x^*$.
\floatsetup[figure]{style=plain,subcapbesideposition=top}

In the following paragraph the causes leading the jets to have different 
spreading are explained by looking at the vorticity fields which, being 
affected by the corners, dictate the evolution of the velocity structures 
previously described. The effect of corners is enlightened in figure 
\ref{OXV_J}, where the positive and negative contours of $\omega_x$ are 
respectively coloured in red and blue, and are overlapped to the contour of
$U=U_{\infty}/2$. The figure clearly shows that in $J_S$, $J_F$ and $J_6$, 
the thin layers of $\omega_x$ produced at the corners, are responsible for 
the rotation and successive stretching of the jets. The formation of 
$\omega_x$ can not be instead observed in $J_C$. At the nozzle exit, in 
correspondence of every corner, a pair of thin layers of $\pm \omega_x$ 
vortices forms. While these are convected downstream, the positive vortices
come up against the negative ones generated by another corner, thus forming
a new couple. This interaction between $\omega_x$ of opposite sign 
stretches the spikes and spreads the jet. For the fractal $J_F$, 
$\pm \omega_x$ of different size are produced at the inlet but, due to the 
low $Re$, the small patches die and do not affect the flow; therefore at a 
certain distance the jet is affected only by the six largest structures and
does not differ from $J_6$.

\begin{figure}[h!]
\begin{center}
\sidesubfloat[]
{\includegraphics[width=0.25\textwidth,angle=0]{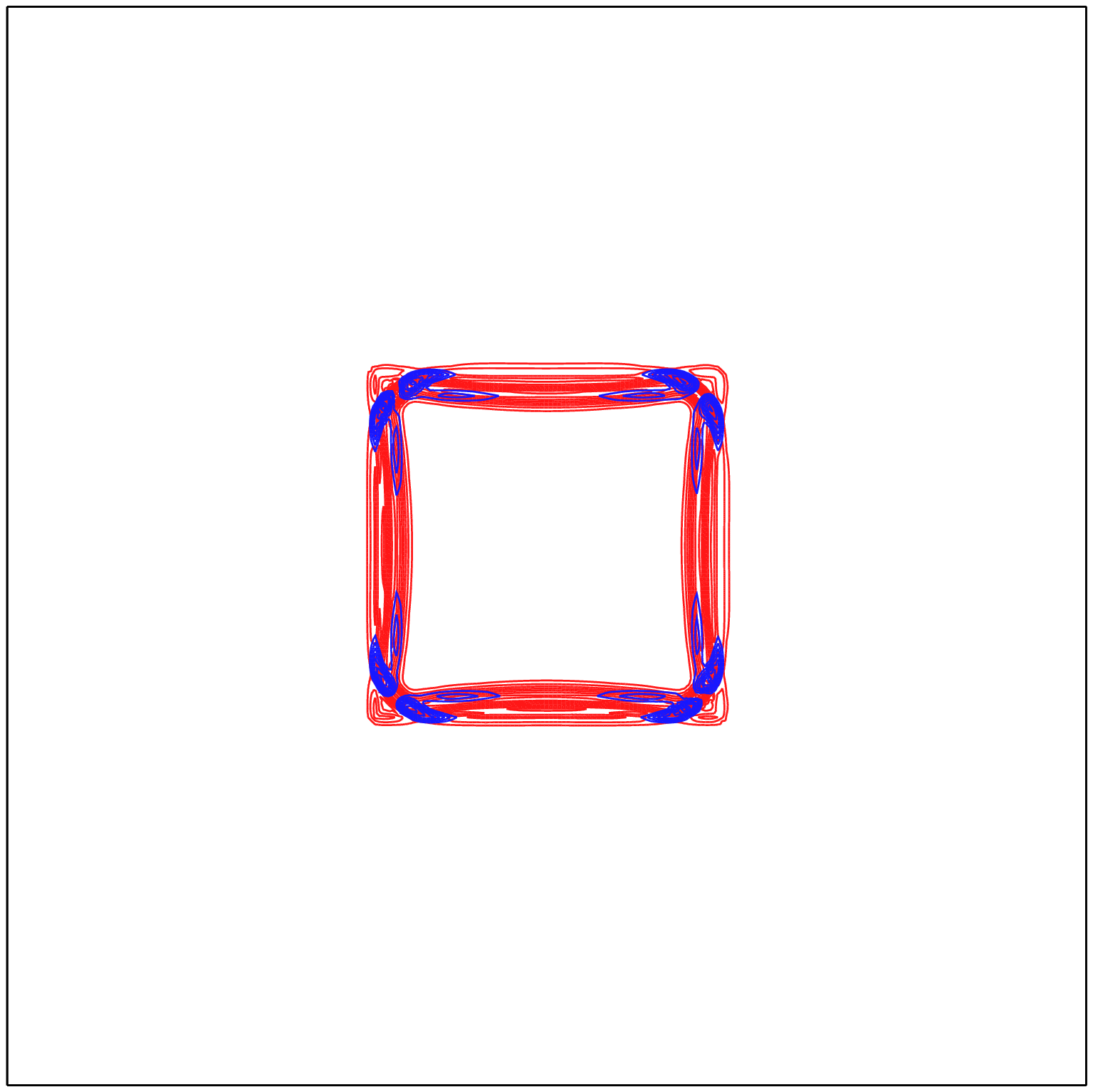}}
\sidesubfloat[]
{\includegraphics[width=0.25\textwidth,angle=0]{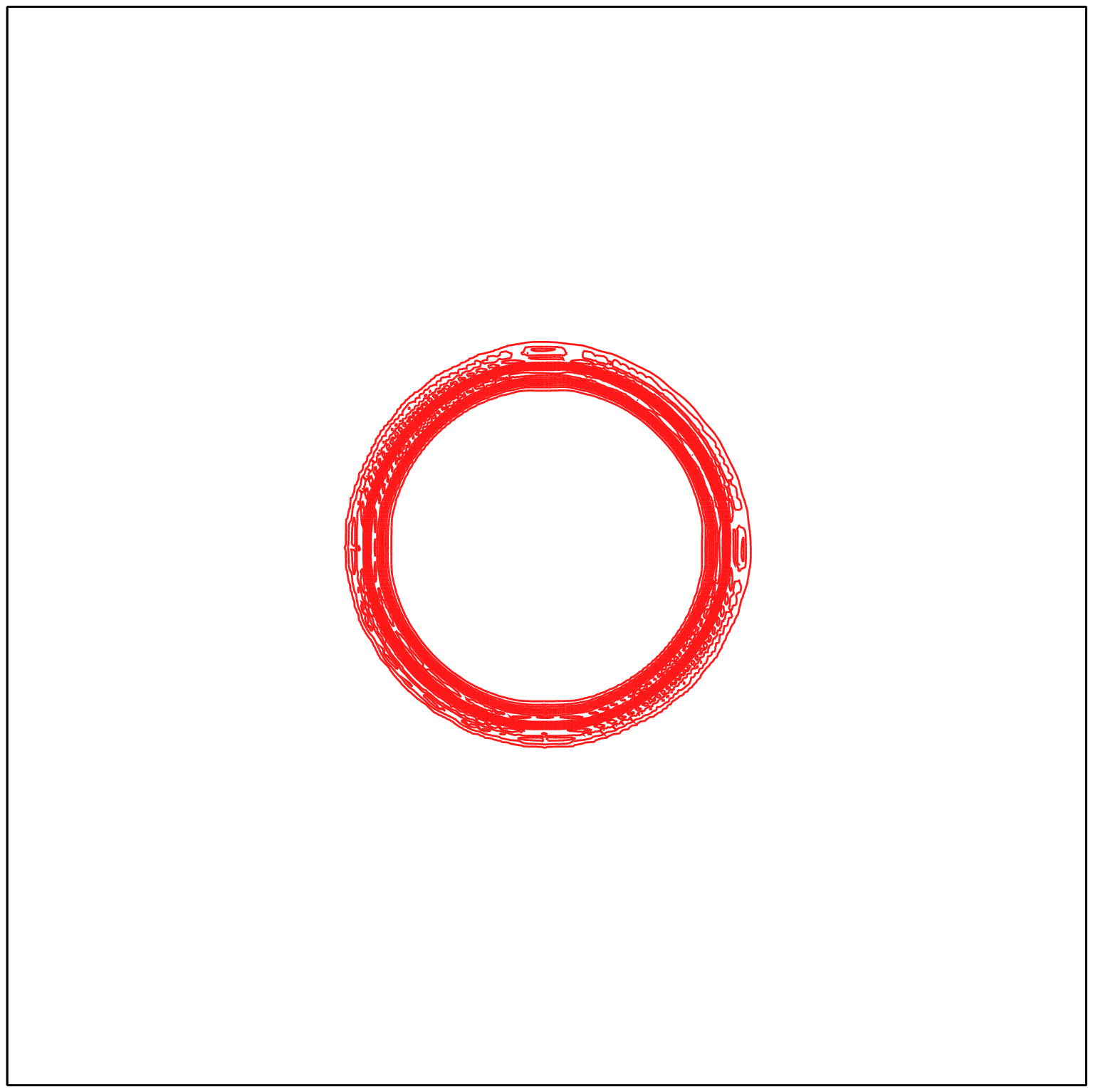}}
\sidesubfloat[]
{\includegraphics[width=0.25\textwidth,angle=0]{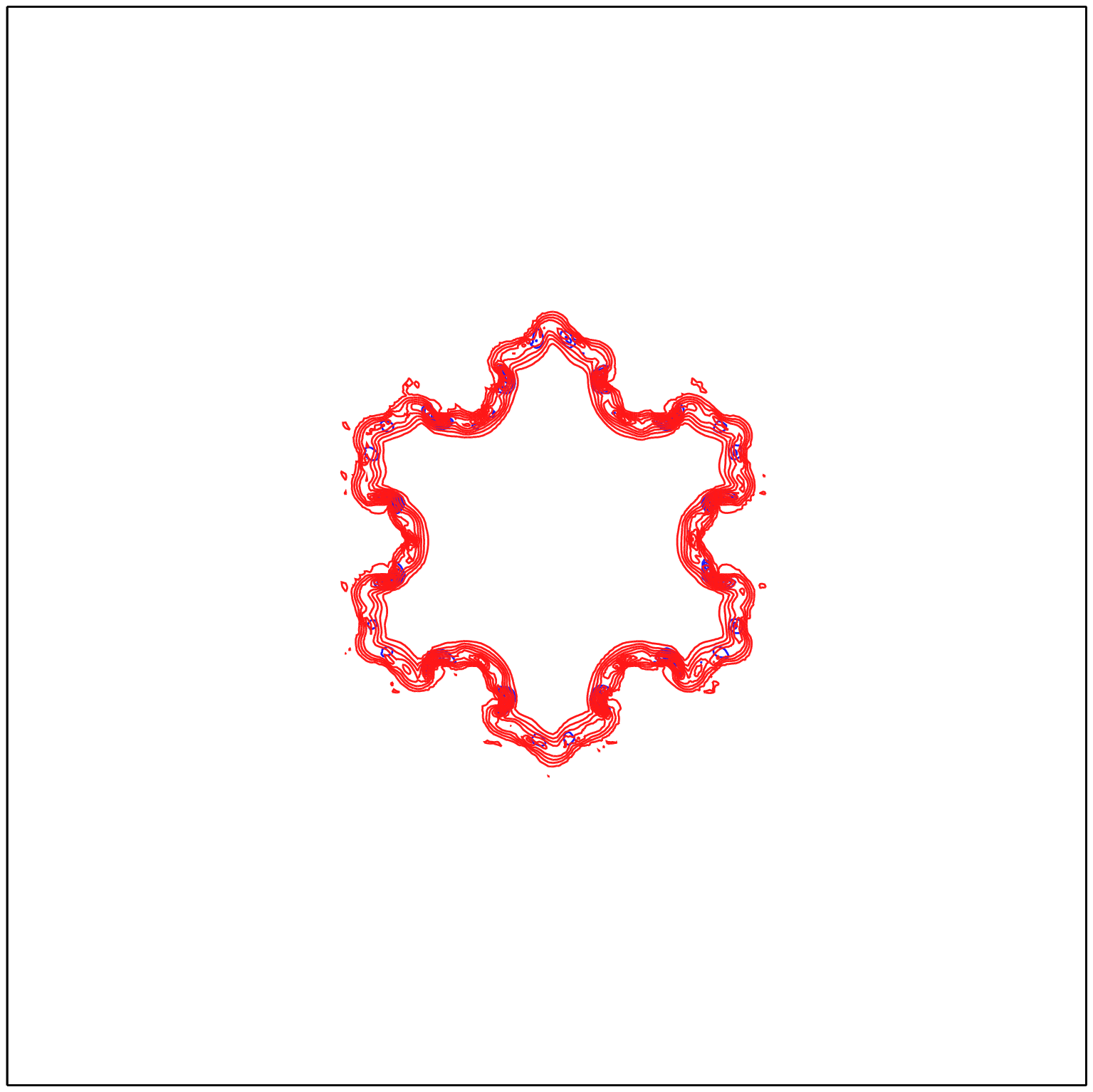}}\\
\sidesubfloat[]
{\includegraphics[width=0.25\textwidth,angle=0]{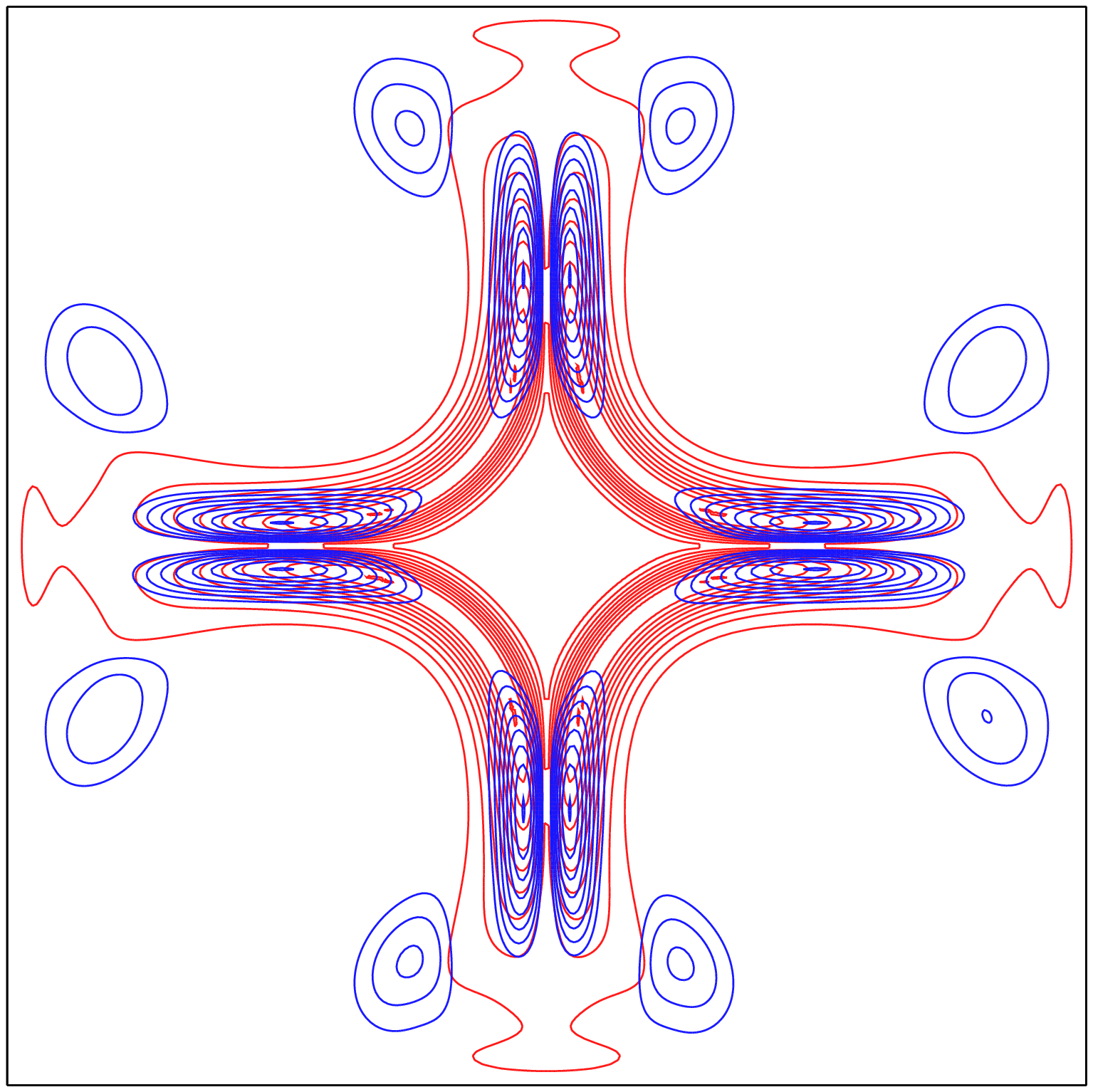}}
\sidesubfloat[]
{\includegraphics[width=0.25\textwidth,angle=0]{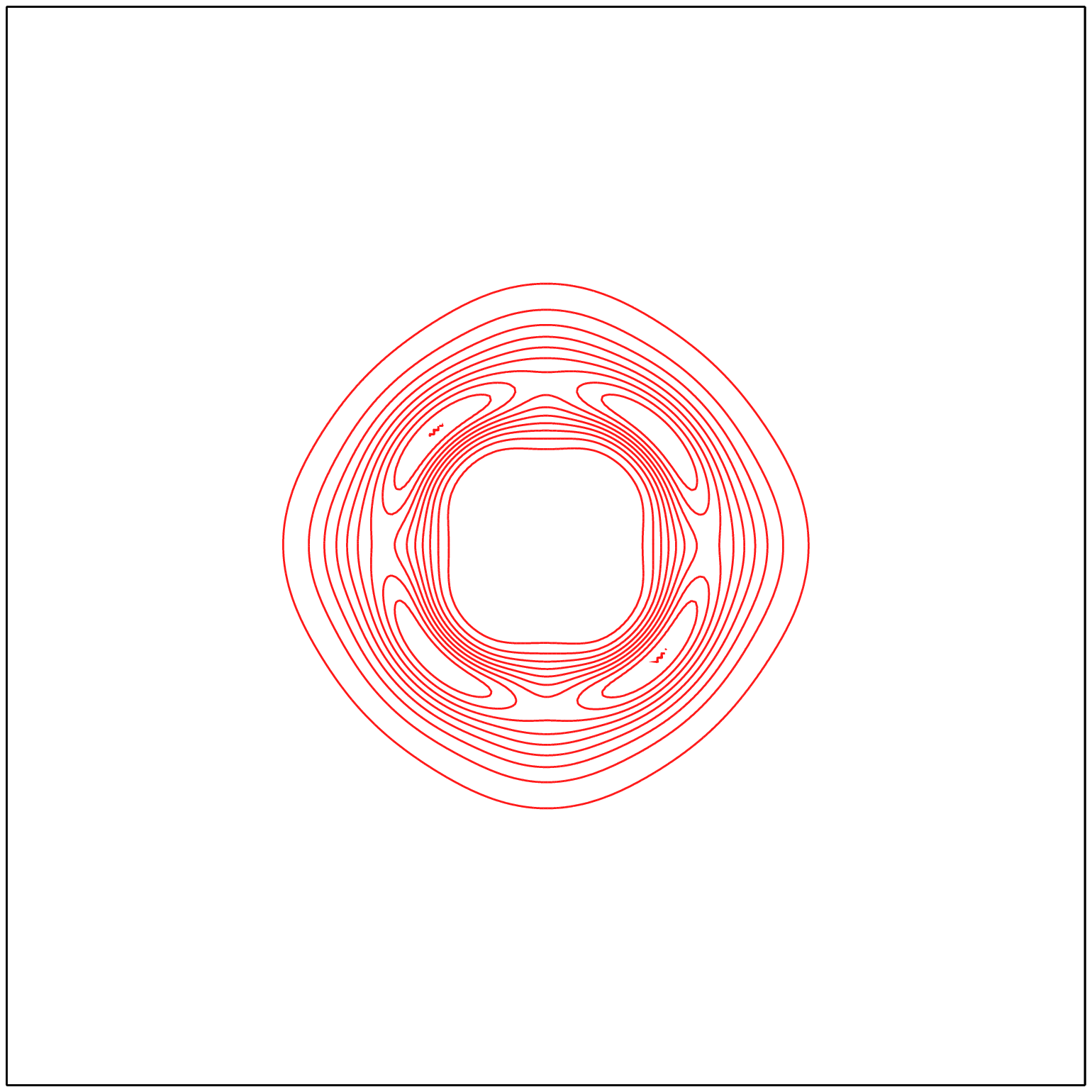}}
\sidesubfloat[]
{\includegraphics[width=0.25\textwidth,angle=0]{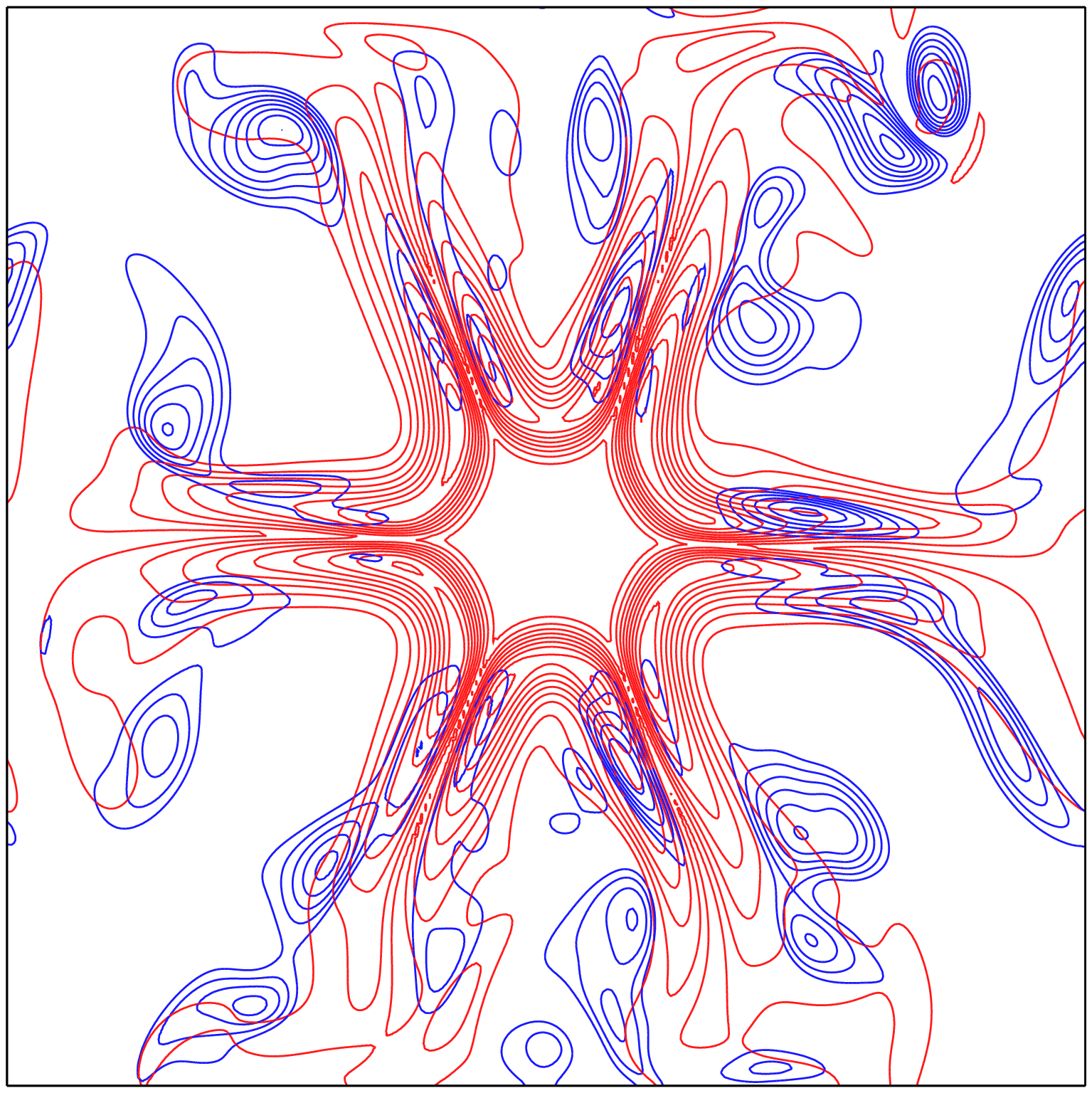}}\\
\caption{Contour plot of $\Omega_{\theta}=\sqrt{\omega_y^2+\omega_z^2}$ 
normalized with its maximum value in the range $[0.1:1]$ with 10 contour 
lines in red for $(a)$ $J_{S}$, $(b)$ $J_{C}$ and $(c)$ $J_{F}$ at $x^*=0$;
$(d)$, $(e)$ and $(f)$ at $x^*=3.18$. The plot is overlapped with the 
contours of $\Omega_x$ normalized with its maximum value in the range 
$[0.3:1]$ with 7 contour lines in blue.}
\label{2Doth}
\end{center}
\end{figure}

In addition to these vortical structures, at the end of the computational 
domain of $J_{F}$ (figure \ref{OXV_J}c) and $J_{6}$ (figure \ref{OXV_J}d), 
other structures with $\pm \omega_x$, that are not connected to the layers
generated at the corners, are visible. These structures are due to the 
vortex dynamics that, through the processes of vortex-stretching and 
vortex-tilting, generates $\omega_x$ from the other vorticity components. 
In fact at the inlet $\omega_y$ and $\omega_z$ are also generated by the 
orifice, forming a corrugated sheet of intense vorticity. Such structures 
for the three jets $J_{C}$, $J_{S}$ and $J_{F}$ are shown in figure 
\ref{2Doth}, where isocontours of 
$\Omega_{\theta}=\sqrt{\omega_y^2+\omega_z^2}$ and 
$\Omega_x=|\omega_x|$, in the plane $yz$ are plotted at two distances 
$x^*$. The quantity $\Omega_{\theta}$ for $J_{C}$ shows a structure with 
the same shape of the vorticity $\omega_{\theta}$ for a circular jet, 
evaluated in cylindrical coordinates \cite{vo1993}. In the figure the 
region with a clustering of the contour lines are those with the highest 
values of $\Omega_{\theta}$, and represents the boundaries of the vortex 
sheet. The increase of the width of this region with $x^*$ indicates the 
jet $J_{C}$ is spreading both towards the center-line and the surroundings.
The other jets present a more complicated pattern at $x^*=3.18$, because 
the jet spreading is greater than for $J_C$, and the fluid transport 
between the jet and its surroundings is enhanced. This fast mechanism, 
triggered by the $\Omega_{x}$ formed at the corners of the nozzle, is the 
cause for the intricate vorticity pattern, where also $\Omega_{x}$ not
generated at the inlet by the orifice is present.
\floatsetup[figure]{style=plain,subcapbesideposition=top}

Figure \ref{bhsf}b reports the evolution of 
$\left\langle \Omega_{\theta} \right\rangle$ and 
$\left\langle \Omega_x \right\rangle$ averaged in the plane $yz$ and in 
time, along the stream wise direction. The values achieved by
$\left\langle \Omega_{\theta} \right\rangle$ are larger than 
$\left\langle \Omega_x \right\rangle$ for any jet, and the plot shows that 
$\left\langle \Omega_{\theta} \right\rangle \propto 1/\theta_{corner}$; in 
fact the highest values are achieved by $J_{F}$ and $J_{6}$ that have a 
smaller $\theta_{corner}$ with respect to $J_S$. The plot shows
that $\left\langle \Omega_x \right\rangle$ is null along the whole domain 
for the circular jet. In the other cases 
$\left\langle \Omega_x \right\rangle$ is approximately constant, until 
$x^* \approx 2.25$. The subsequent growth is due to the vorticity dynamics 
that produces $\left\langle \Omega_x \right\rangle$ from 
$\left\langle \Omega_{\theta} \right\rangle$ by vortex stretching 
and vortex tilting. The high $\left\langle \Omega_{\theta} \right\rangle$ 
is characterized by a first region where this quantity decreases, followed 
by a second region where it grows linearly. At the end the decrease is 
mainly due to the viscosity. The slope and the length of the linear region 
depend on the geometry of the orifice, in fact $a_{J_{6}}=0.14$ and 
$a_{J_{F}}=0.14$, while $a_{J_{S}}$ is $0.10$ and the value of 
$\left\langle \Omega_{\theta} \right\rangle$ for $J_{C}$ remains 
approximately constant ($a_{J_{C}}=0.01$).

\vspace{-0.2cm}
\section{Conclusion}
The present DNS of the jets produced by nozzles of complex geometries at 
$Re_{D_e}=565$ allowed to visualize the complex vortex dynamics, to 
increase the understanding of the jet evolution, in order to enhance the 
entrainment and the mixing. 
The jets generated from nozzles with sharp corners evolve differently from
circular jets, that maintain an axisymmetric circular shape, due to the 
axis switching of the jet. This rotation is produced at the nozzle and is 
initiated by the vorticity self-induction process. The rotation of the jet
has been related to the angular width of the corners  
($\theta_{rotation}=\theta_{corner}/2$) and, from the data in literature, 
the relationship between $\theta_{rotation}$ and $\theta_{corner}$ seems to
be valid for nozzles with aspect ratio $1$. The velocity at which the 
rotation occurs can be measured by the growth rates of the $s-$ and 
$l-$diagonals and it has been shown that, while  $r_{1/2}^l$ is similar
for all the nozzles with corners, $r_{1/2}^s$ is strongly affected by 
the shape. The growth of $r_{1/2}^s$ is associated with the stretching 
of the corrugations, in fact in $J_F$ and $J_6$ $r_{1/2}^s$ grows faster,
the corners undergo a faster stretching, and this process contributes to 
enhance the spreading. The deformation of the jet is linked to the vortex 
dynamics because patches of $\pm \omega_x$ are formed at the corners. These
thin layers are convected downstream and in turn deform the jet. The spikes
of the jet are stretched by these vortices and consequently the structure 
is spread more rapidly, higher is the strength of the $\omega_x$ patches.
As a result the fluid transport between the jet and its surroundings is 
enhanced, so as the mixedness. The low Reynolds number assumption 
guarantees that the flow remains laminar in the near-field and, in such 
condition, the wider range of scales generated by the fractal nozzle, does 
not affect the flow, therefore the evolution of $J_F$ and $J_6$ does not 
show large differences.

\input{referenc}

\end{document}

%% file: referenc.tex
%
%
%